\documentclass[
  amsmath,amssymb,
  aps,
  pre,
  superscriptaddress
]{revtex4-2}

\usepackage{graphicx}
\usepackage{dcolumn}
\usepackage{bm}

\usepackage{xcolor}

\begin{document}

\title{Intermittency in Shell Models of Turbulent Cascades:\\
from Single-Branch to Multi-Branch}

\author{Flavio Tuteri}
\email{flavio.tuteri@phys.ens.psl.eu}
\affiliation{Laboratoire de Physique de l'\'Ecole normale sup\'erieure, ENS, Universit\'e PSL, CNRS, Sorbonne Universit\'e, Universit\'e Paris Cit\'e, F-75005 Paris, France}

\author{Sergio Chibbaro}
\affiliation{LISN, AO team, B\^atiment 660, Universit\'e Paris-Saclay, CNRS, 91405 Orsay Cedex, France}
\affiliation{Inria Saclay, TAU team, B\^atiment 660, Universit\'e Paris-Saclay, 91405 Orsay Cedex, France}

\author{Alexandros Alexakis}
\affiliation{Laboratoire de Physique de l'\'Ecole normale sup\'erieure, ENS, Universit\'e PSL, CNRS, Sorbonne Universit\'e, Universit\'e Paris Cit\'e, F-75005 Paris, France}

\date{\today}

% --------------------------------------------------------------------------- |

\begin{abstract}
Intermittency is one of the central features of turbulent transfer: the multi-scale energy cascade is mediated by rare and intense fluctuations.
We investigate this phenomenon in a multi-branch shell model, which combines quasi-local triadic nonlinear interactions with a branching structure that mimics the growth of degrees of freedom toward small scales.
Comparison with the standard Sabra model shows that branching enhances intermittency, as measured by anomalous scaling exponents of energy-flux structure functions.
We further use multiplier statistics and large deviation estimates to characterize the multiplicative nature of the cascade.
Our results suggest that reduced descriptions of turbulent intermittency should retain both nonlinear dynamics and geometrical organization. Implications on Navier-Stokes turbulence are discussed.
\end{abstract}

%\keywords{}

\maketitle

% --------------------------------------------------------------------------- |

\section{\label{sec1} Introduction}

Intermittency remains the central issue in the theory of three-dimensional turbulence~\cite{alexakis2018cascades}: rare and intense fluctuations spoil the scale-invariant picture suggested by the inviscid limit.
A turbulent flow is described by the forced and viscous Navier-Stokes equations,
\begin{align}
    \partial_t\boldsymbol{v}+\boldsymbol{v}\cdot\nabla\boldsymbol{v}
    &=
    -\nabla P+\nu\nabla^2\boldsymbol{v}+\boldsymbol{f},
    \label{NSR1}\\
    \nabla\cdot\boldsymbol{v}&=0,
    \label{NSR2}
\end{align}
where $\boldsymbol{v}$ is the velocity field, $P$ is the pressure, $\nu$ is the viscosity, and $\boldsymbol{f}$ is an external forcing.
As a nonlinear system of partial differential equations, any analytical treatment is challenging.
For this reason, simplified models that retain the symmetries and inviscid invariants of the Navier-Stokes equations have long been used to understand and quantify the phenomenon of intermittency in a reduced setup.
Among the most popular models used for this purpose are the so-called shell models, which are inspired by the wavenumber-space form~\cite{frisch1995turbulence} of Eqs.~\eqref{NSR1}-\eqref{NSR2}:
\begin{align}
    \big[\partial_t+\nu\,k^2\big]\hat{v}_j(\mathbf{k},t)=
    -i\sum_{\mathbf{p}\,+\,\mathbf{q}\,=\,\mathbf{k}}\mathbb{P}_{jh}(\mathbf{k})\,k_r\,\hat{v}_r(\mathbf{q},t)\,\hat{v}_h(\mathbf{p},t)+\hat{f}_j(\mathbf{k},t),
    \label{NSF}
\end{align}
where $\mathbb{P}$ is the Leray projector onto divergence-free fields.
The incompressible nonlinearity is nonlocal and couples different scales.
Shell models reduce Eq.~\eqref{NSF} by replacing all Fourier modes with wavenumber $k$ in the logarithmic shell $k_0\lambda^l \le k < k_0\lambda^{l+1}$ by a single complex amplitude $u_l$.
The variables $u_l$ then evolve according to nonlinear equations designed to reproduce the triadic structure of Eq.~\eqref{NSF}.
Furthermore, only quasi-local interactions between shells are retained, in the spirit of the cascade-locality arguments~\cite{eyink2005}.
Despite this drastic simplification, shell models reproduce the energy cascade of three-dimensional turbulence and display intermittent statistics~\cite{biferale2003}.
However, this corresponds to a temporal form of intermittency, since extreme events occur over short time intervals, while all spatial information has been averaged out.
Such temporal fluctuations are also present in real turbulence, but they coexist with strongly intermittent spatial structures that standard shell models cannot represent. \medskip

One possible way to quantify spatial intermittency is through the local energy transfer across scales.
At a given length scale $\ell$, the large-scale component is obtained by averaging with a filter $G$:
\begin{eqnarray}
    {\overline{v_j}}^\ell(\mathbf{x},t)=\int G^\ell(\mathbf{r})\,v_j(\mathbf{x}-\mathbf{r},t)\,d\mathbf{r},
\end{eqnarray}
where $G^\ell(\mathbf{r})=\ell^{-3}G(\mathbf{r}/\ell)$. This coarse-graining framework~\cite{germano1992} provides a definition of the local energy flux around $\mathbf{x}$, at time $t$, and across the scale $\ell$:
\begin{eqnarray}
\label{localFlux}
    \Pi^\ell(\mathbf{x},t)=-\big[\overline{v_i v_j}^\ell-{\overline{v_i}}^\ell\,{\overline{v_j}}^\ell\big]\,\partial_j{\overline{v_i}}^\ell.
\end{eqnarray}
Moments of $\Pi^\ell$ provide a measure of intermittency. In the inertial range, a power-law behavior is observed
\begin{eqnarray}
    \langle\lvert\Pi^\ell\rvert^{q/3}\rangle_{\mathbf{x},t}\sim\ell^{\,\xi_{q/3}}
\end{eqnarray}
for which $\xi_{q/3}= 0$ for all $q$ corresponds to a non-intermittent (scale invariant) behavior of $\Pi^\ell$. However, measurements indicate that $\xi_{q/3}\neq 0$ for $q\neq3$, reflecting the long tails in the distributions of $\Pi^\ell$~\cite{chen2003,alexakis2020local,tuteri2026fluctuations}.
This result also holds for single field realizations, with only spatial averaging and no time averaging.
This points to a fundamental difference between intermittency in shell models and in Navier-Stokes.
Turbulent flows display both temporal and spatial fluctuations, whereas shell models contain no geometrical information and generate intermittency through temporal variability alone, pointing to a purely dynamical source of fluctuations~\cite{aumaitre2024}. \medskip

A way to incorporate this missing geometrical information into shell modeling is to introduce a $p$-adic branching structure, in which each mode at scale $\ell$ is connected to $p$ modes at the smaller scale $\ell/\lambda$.
As a result, the number of degrees of freedom grows exponentially toward small scales.
Rather than representing an average over Fourier modes in a logarithmic shell, each variable represents a velocity structure at a given scale and location, thereby retaining both temporal and spatial variations.
The increasing number of modes follows the Navier-Stokes scaling and allows the thermalized state of the system to be correctly captured~\cite{tuteri2026dual}, unlike in classical shell models.
This class of multi-branch shell models, however, has received comparatively little attention~\cite{zimin1981hierarchic,aurell1994,benzi1997,aurell1997}.
An astonishing result recently shown for this class of models is that they allow for the construction of a family of exact, finite-flux, time-independent solutions that display intermittency~\cite{ajzner2023,tuteri2026}.
Thus, in contrast with classical shell models, where intermittency is generated by the dynamics, these solutions are time independent and their intermittency originates solely from the geometrical structure, namely from spatial variability~\cite{ajzner2023,tuteri2026}.\medskip

In real turbulence, both temporal and spatial sources of intermittency are present.
In this work, we study intermittency in a dynamically evolving multi-branch shell model and compare its statistics with those of classical single-branch models.
Moreover, we extend the multiplier and large deviation approach used to describe intermittency in classical shell models~\cite{mailybaev2013} to this setting, highlighting the role of multiple paths in energy transfer.
Implications for Navier-Stokes turbulence are discussed.

% --------------------------------------------------------------------------- |

\section{\label{sec2} Shell Models}

First we introduce the shell models studied in this work, starting from the standard single-branch case and then moving to its multi-branch extension.

\subsection{\label{secI2} Single-branch Models}

The classical Sabra model~\cite{lvov1998sabra} reads
\begin{align}
    \big[d_t+\nu\,{k_l}^2\big]u_l=i\,k_l\big[
    a\,\lambda     \,{u_{l+1}}^\ast u_{l+2}+
    b\,              {u_{l-1}}^\ast u_{l+1}-
    c\,\lambda^{-1}\, u_{l-2}       u_{l-1}
    \big]
    +f_l
    \label{SabraSingle}
\end{align}
where 
\begin{equation}
    a=1,\quad b=-(1-\lambda^{-1}),\quad c=-\lambda^{-1}.
\label{model_parameters}
\end{equation}
Here $u_l(t)\in\mathbb{C}$ is the complex variable associated with the wavenumber $k_l=k_0\lambda^l$ (or equivalently the scale $\ell_l=\ell_0\lambda^{-l}=k_l^{-1}$), where $\lambda>1$ is the fixed inter-shell ratio.
For zero viscosity and zero forcing, it conserves energy $E=\frac{1}{2}\sum_l|u_l|^2$ and a helicity-like quantity $H=\frac{1}{2}\sum_l (-1)^l k_l |u_l|^2$. 
The energy flux across a level $l$~\cite{pisarenko1993} is
\begin{eqnarray}
    \Pi_l(t)=k_l\,\mathrm{Im}\big[\lambda\,a\,u_l u_{l+1} {u_{l+2}}^\ast-c\,u_{l-1}u_{l}{u_{l+1}}^\ast\big].
\end{eqnarray}
Compared with Eq.~\eqref{localFlux}, this object is simpler because the scale dependence is discrete and it carries no spatial dependence.
In the present work we will focus on the statistics of $\Pi_l$ and compare them with those of the dyadic-tree Sabra model described below.

% --------------------------------------------------------------------------- |

\subsection{\label{secII2} Multi-branch Models}

\begin{figure}
\includegraphics[width=.9\columnwidth]{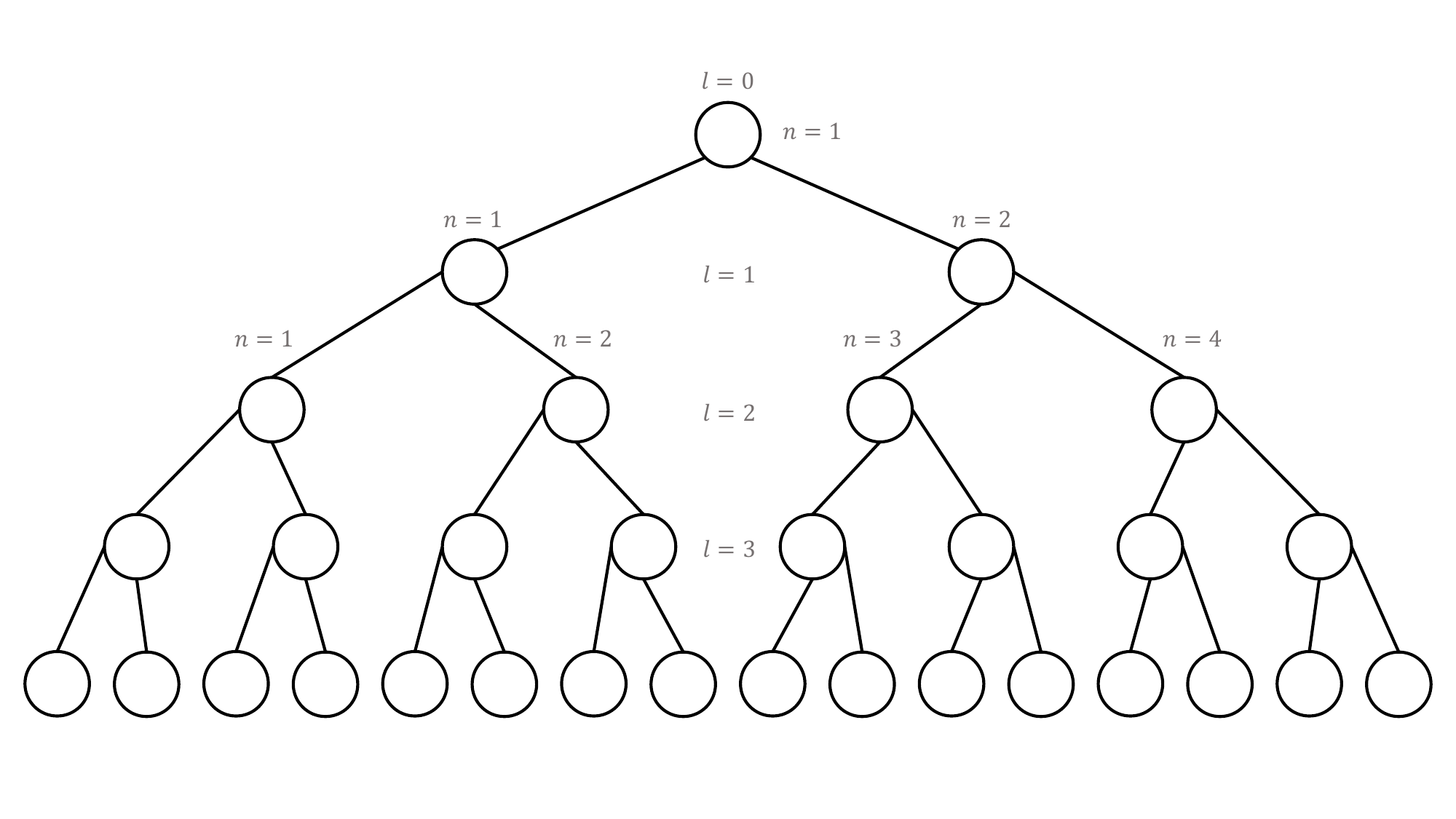}
\caption{\label{fig:p-adic} Dyadic tree structure for the multi-branch model.}
\end{figure}

Fix the inter-shell ratio $\lambda$. 
A general multi-branch shell model follows a $p$-adic tree structure, shown in Fig.~\ref{fig:p-adic} for the dyadic case.
The system state consists of complex dynamical variables $u_{l,n}$ indexed by the $p$-adic tree level and position:
\begin{eqnarray}
    (l,n)\quad\text{with}\quad 
    l\in\big\{0,\ldots,L  \big\},\;
    n\in\big\{1,\ldots,p^l\big\}.
\end{eqnarray}
Each node $(l,n)$, except the `{\it root}' $(0,1)$, has a single parent $(l-1,\lceil n/p\rceil)$, where $\lceil x \rceil =\min\{m\in\mathbb{Z}\mid m\ge x\}$ is the ceiling.
Each node $(l,n)$, except the `{\it leaves}' $(L,n)$, has $p$ children, namely $(l+1,p(n-1)+1),\ldots,(l+1,pn)$.
The Sabra coupling on this topology is written as
\begin{eqnarray}
    \bigg[\frac{d}{dt}+\nu{k_l}^{2}\bigg]u_{l,n}=i N_{l,n}[u]+f_{l,n},
    \label{multiSabra}
\end{eqnarray}
where $\nu$ is the viscosity, $f$ the large-scale forcing, and the nonlinear coupling is given by
\begin{alignat}{3}
    N_{l,n}=k_l\bigg[&a\,\lambda\,\frac{1}{p}\sum_{h=0}^{p-1}\bigg( &&{u_{l+1,pn-h}              }^\ast \,\frac{1}{p}\sum_{m=0}^{p-1} &&u_{l+2,p(pn-h)-m}\bigg)\nonumber\\
                  +\,&b                                             &&{u_{l-1,\lceil n/p  \rceil}}^\ast \,\frac{1}{p}\sum_{m=0}^{p-1} &&u_{l+1,pn-m}            \nonumber\\
                  -\,&c\,\lambda^{-1}                               && u_{l-2,\lceil n/p^2\rceil}                               &&u_{l-1,\lceil n/p\rceil}\bigg],
\label{nonlinear}
\end{alignat}
with $u_{-2,\bullet}=u_{-1,\bullet}=0=u_{L+1,\bullet}=u_{L+2,\bullet}$.
All the global observables are defined in accordance with the interpretation of $l$ as a scale index and $n$ as a spatial index.
This construction generalizes the Sabra model: each triad involving the same levels and compatible with the genealogy is taken into account with the same weight and averaged.
Thus, single-branch solutions $u_l$ of Eq.~\eqref{SabraSingle} are the homogeneous solutions of Eq.~\eqref{multiSabra}, with $u_{l,n}=u_l$ for all $n$.
The total energy is defined as a volume-weighted sum of the local energy density,
\begin{equation}
    E=\frac{1}{2}\sum_{l=0}^{L}\frac{1}{{k_l}^D}\sum_{n=1}^{p^l}\lvert u_{l,n}\rvert^2,
\end{equation}
where $D$ denotes the effective spatial dimension and is related to $\lambda$ by the relation
\begin{equation}
    D^{-1}=\log_p\lambda.
\label{spaDim}
\end{equation}
Equation~\eqref{spaDim} ensures energy conservation and has a clear interpretation: for a space-filling cascade, the number of substructures $p$ equals the volume reduction factor ${k_{l+1}}^D / {k_{l}}^D=\lambda^D$ between successive scales so that the volume $V_{l,n}={k_l}^{-D}$ of a structure $(l,n)$ equals the sum of the $p$ volumes of the sub-structures at scale $l+1$ linked to $(l,n)$, i.e. $V_{l,n}
%=\sum_{n=1}^p V_{l+1,n}
=pV_{l+1,n}$.
The inviscid and unforced dynamics conserves the total energy and the helicity-like quantity
\begin{equation}
    H=\sum_{l=0}^{L}(-1)^l{k_l}^{1-D}\sum_{n=1}^{p^l}\lvert u_{l,n}\rvert^2,
\end{equation}
given Eq.~\eqref{spaDim} and provided that $a,b,c$ satisfy Eq.~\eqref{model_parameters}.
The local energy transfer from $(l-1,\lceil n/p\rceil)$ to $(l,n)$ is defined as the rate of change of the energy contained in $(l,n)$ and in all descendant nodes, weighted by the volume $V_{l,n}$.
In other words, $\Pi_{l,n}$ is an energy flux density, as in the definition of the
local energy flux in Eq.~\eqref{localFlux} for the Navier Stokes turbulence.
This leads to
%Working with the densities,
\begin{align}
    \Pi_{l,n}&={k_l}^D\operatorname{Im}\!\bigg[\sum_{j=l}^{L}\frac{1}{{k_j}^D}\sum_{m=p^{j-l}(n-1)+1}^{p^{j-l}n} u_{j,m}N_{j,m}[u]^\ast\bigg] \\
             &=k_{l}\operatorname{Im}\!\bigg[-a\,u_{l-1,\lceil n/p\rceil}u_{l,n}\,\frac{1}{p}\sum_{m=0}^{p-1}{u_{l+1,pn-m}}^\ast 
             +c\,\lambda^{-1}\, u_{l-2,\lceil n/p^2\rceil}u_{l-1,\lceil n/p\rceil}{u_{l,n}}^\ast\bigg].
\label{local_flux}
\end{align}
With this definition, the average over time of $\Pi_{l,n}$ in the inertial range is equal to the average of $\Pi_{l+1,m}$.
%Expression~\eqref{local_flux} follows from energy conservation.
\medskip

In the following, we focus on the dyadic case $p=2$.

% --------------------------------------------------------------------------- |

\section{Energy Cascades}

\begin{figure}[h]
\includegraphics[width=.9\columnwidth]{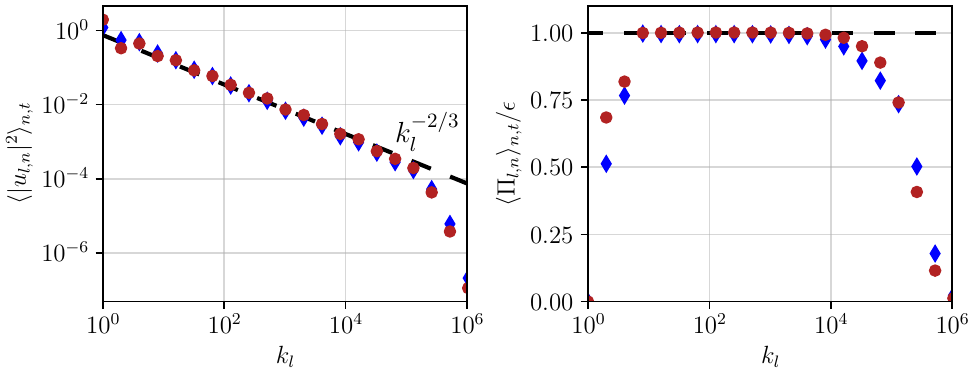}
\caption{\label{figSpectra}
Energy spectrum on the left and mean energy transfer, normalized by the mean energy injection rate, on the right. Circles correspond to the multi-branch model, while diamonds correspond to the single-branch model.
}
\end{figure}
\begin{figure}[h]
\includegraphics[width=.9\columnwidth]{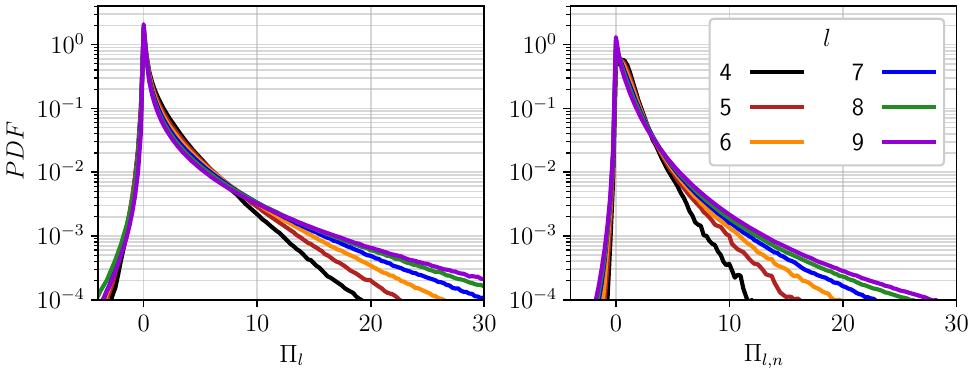}
\caption{\label{figFpdfs} Probability density functions of the energy flux at different scale indices $l$ for the single-branch model, left, with statistics accumulated in time, and for the multi-branch model, right, with statistics accumulated in both time and space. 
}
\end{figure}
We perform simulations of the single-branch and dyadic $p=2$ shell models with parameters $\lambda=2$, $\nu=10^{-8}$ and $L=24$. 
The first three levels are forced with imposed energy injection rate $\epsilon=1$.
The results examined in what follows are statistically averaged in the stationary state.
The resulting spectra are compatible with a three-dimensional turbulent cascade, as shown in Fig.~\ref{figSpectra}.
Both models display a near-Kolmogorov energy spectrum $E\sim {k_l}^{-2/3}$ and a constant energy flux in the inertial range.

\subsection{Measures of Intermittency}
\begin{figure}[h]
\includegraphics[width=.9\columnwidth]{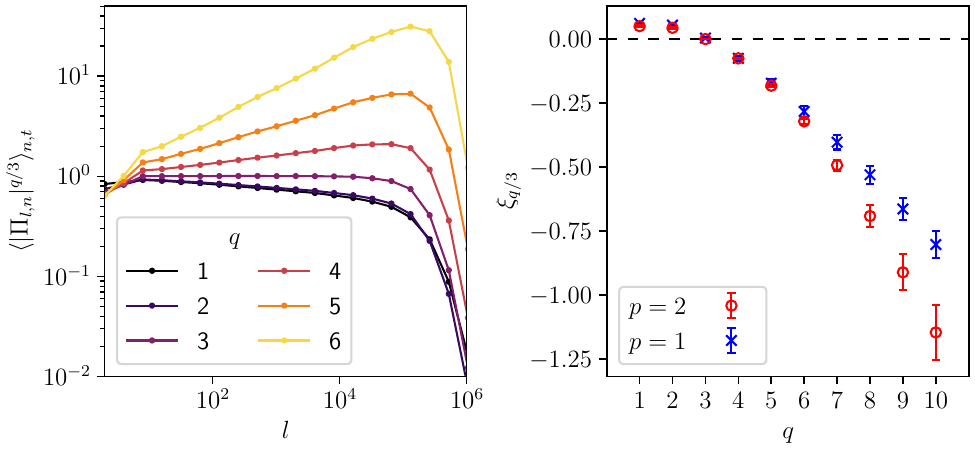}
\caption{\label{figScaling} Structure functions $\langle |\Pi_{l,n}|^{q/3} \rangle_{n,t}$ for the $2$-adic model, left, and the corresponding scaling exponents, right; the right panel also compares them with those of the classical $1$-adic model.}
\end{figure}
To quantify intermittency, we focus on the statistics of the energy fluxes $\Pi_l$ and $\Pi_{l,n}$, measuring their distributions and their moments $\langle |\Pi_l|^m \rangle_t$ and $\langle |\Pi_{l,n}|^m \rangle_{n,t}$. The angular brackets denote averages over the indicated indices.
In Fig.~\ref{figFpdfs} we show the probability density functions of $\Pi_{l}$ and $\Pi_{l,n}$ for the $1$-adic and $2$-adic models for different values of the scale index $l$. The PDFs are not scale independent. On the contrary, the larger $l$, the longer the positive tails in the distribution. This implies that higher moments of $\Pi_{l}$ and $\Pi_{l,n}$ are increasingly affected by rare events.
\medskip

The left panel of Fig.~\ref{figScaling} shows the structure functions for the $2$-adic model.
We fit the measured structure functions with power laws in the inertial range,
\begin{eqnarray}
    \langle\lvert\Pi_{l,n}\rvert^{q/3}\rangle_{n,t}\sim {k_l}^{-\xi_{q/3}}.
\end{eqnarray}
The resulting power-law exponents are shown in the right panel of the same figure.
They are compared with those measured from the single-branch model.
We use the statistical framework proposed in a recent work~\cite{deWit2024} to accurately estimate mean exponents and errors. We refer to the original paper for the details.
While the scaling exponents are indistinguishable for lower moments, up to $q=4$, the multi-branch shell model turns out to be considerably more intermittent, as indicated by the larger deviations of the higher-order exponents from the scale-invariant prediction $\xi_{q/3}=0$.
It is worth noting that the error bars are computed in such a way that the estimation is conservative. 
Thus the results may be considered significant even for the highest order.

% --------------------------------------------------------------------------- |

\section{Sources of Intermittency}

\begin{figure}[h]                                                
\includegraphics[width=.9\columnwidth]{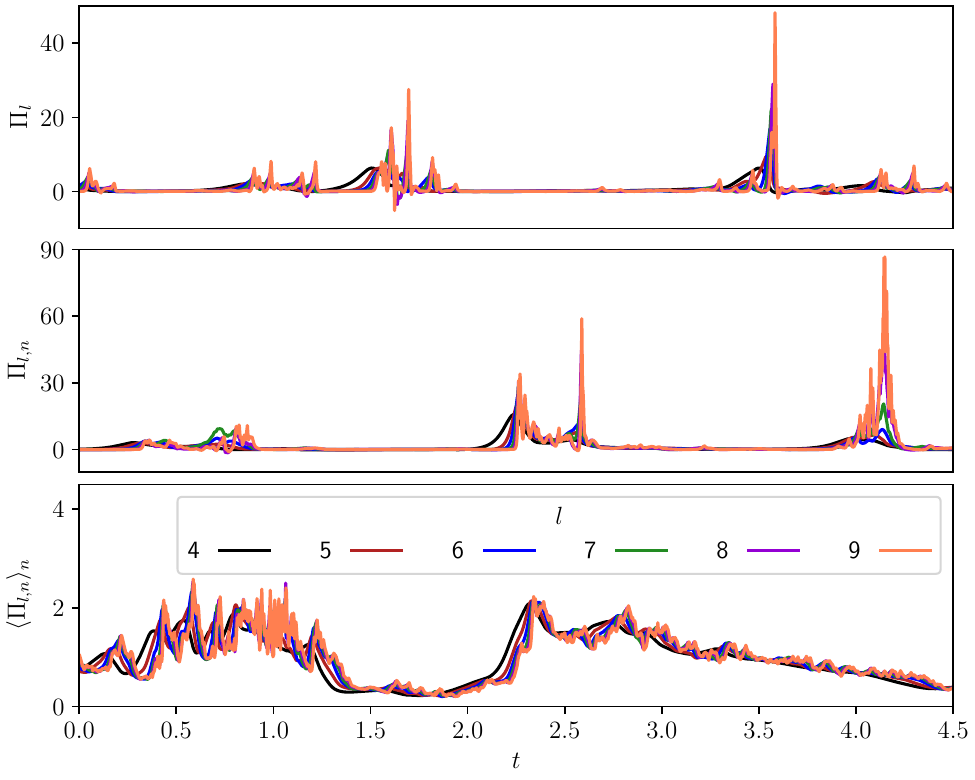}
\caption{\label{figFlux} Energy flux through scales. Comparison between the classical $1$-adic flux $\Pi_l$ (top), the volume-normalized flux $\Pi_{l,n}$ of the $2$-adic model along a walk (middle), and the space-averaged $2$-adic flux $\langle\Pi_{l,n}\rangle_n$ (bottom).}
\end{figure}
\begin{figure} 
\includegraphics[width=.9\columnwidth]{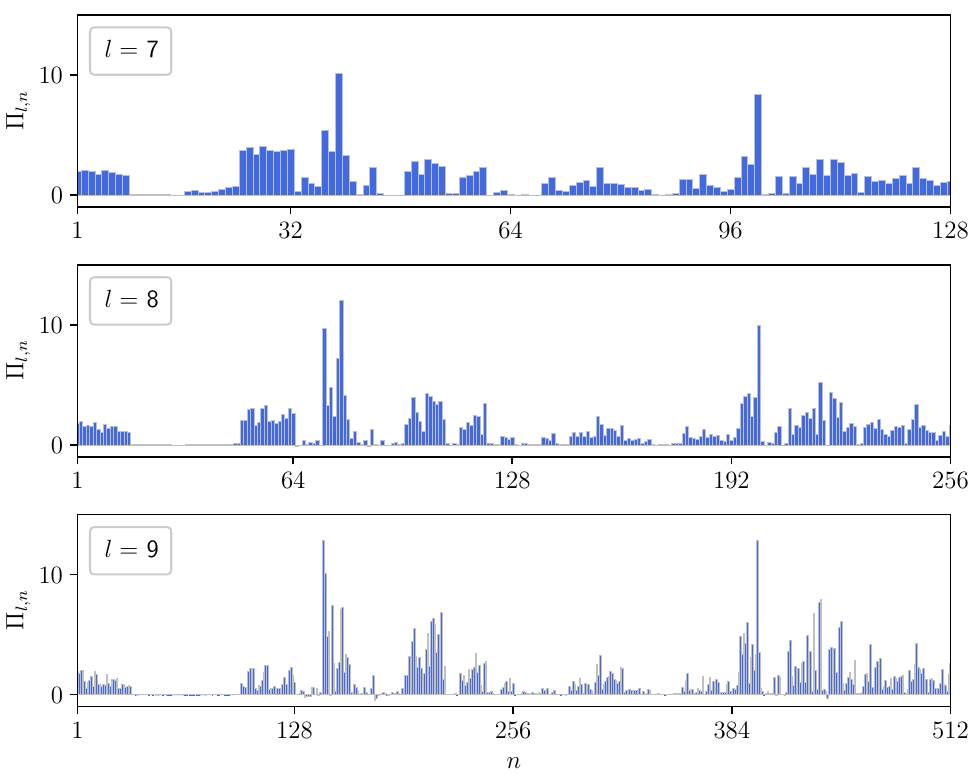}
\caption{\label{fig:n} Multi-branch energy flux as a function of the spatial index $n$ at fixed scale, at a given time.}
\end{figure}                                               %
To understand the deviations from scale-invariant behavior, it is useful to look at representative signals of the energy flux.
First we focus on the dynamics in time.
The top panel of Fig.~\ref{figFlux} shows an example of $\Pi_l(t)$ as a function of time for different values of $l$ inside the inertial range, for a single-branch model realization.
The middle panel shows $\Pi_{l,n}(t)$ for a $2$-adic model, at the same values of $l$, along a randomly chosen path $n=n_l$ on the tree.
Finally, the bottom panel shows $\langle \Pi_{l,n}\rangle_n(t)$, which is the total flux across level $l$.
Strong burst events are present in the top two panels, whereas the bottom panel displays weaker fluctuations, with order-one variations around the mean value in time.
This happens because averaging over $n$ at each level $l$ smoothens out the intense fluctuations observed at individual nodes $(l,n)$.
This behavior is similar to the local energy flux $\Pi^\ell(\mathbf{x},t)$ in Navier-Stokes turbulence: its volume average shows no strong fluctuations, while local fluctuations can be orders of magnitude larger than the mean value~\cite{alexakis2020local,tuteri2026fluctuations}.
It is therefore important to distinguish $\langle {\Pi_{l,n}}^m \rangle_{n,t}$ from $\langle {\langle \Pi_{l,n} \rangle_n}^m\rangle_t$: the former is strongly intermittent, whereas the latter shows no sign of intermittency.\medskip

Then we can look at the impact of space variation.
Figure~\ref{fig:n} shows the local energy flux $\Pi_{l,n}$ obtained from the multi-branch model at one instant of time, as a function of $n$ and for different levels, in the inertial range.
Although all levels have the same time-averaged mean flux, stronger variations are observed as $l$ is increased.\medskip

From these qualitative observations, we can already infer that intermittency in the multi-branch model has two distinct origins: temporal fluctuations along each branch and spatial fluctuations across branches. By contrast, the single-branch model retains only the temporal source of intermittency. These two contributions must now be understood and quantified separately.

% --------------------------------------------------------------------------- |

\subsection{Dynamics}

\begin{figure}[h]
\includegraphics[width=.9\columnwidth]{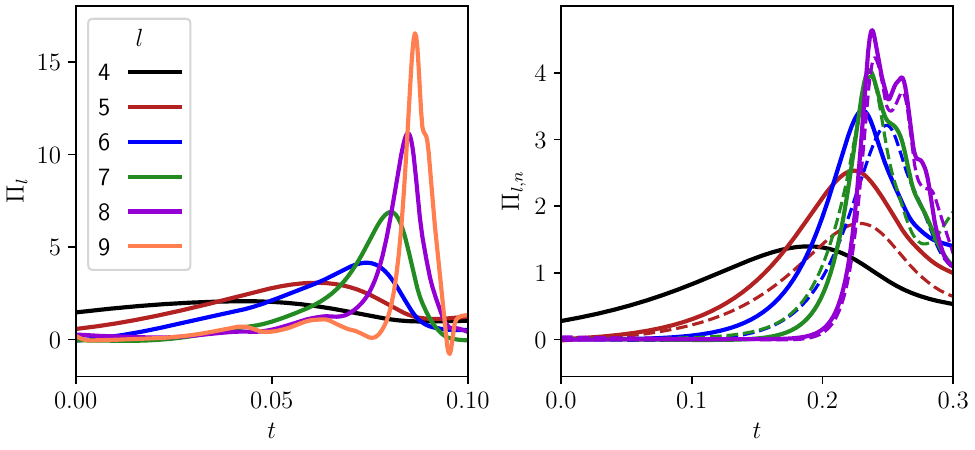}
\caption{\label{figVis} Energy flux through scales focusing on a single burst. Comparison between the single-branch model (left) and a descending path in the multi-branch model (right).}
\end{figure}
Further insight can be gained by zooming in on flux bursts.
For the single-branch model, the bursts increase in amplitude and decrease in duration as $l$ becomes larger.
This is better seen in the zoom of one such event shown in the left panel of Fig.~\ref{figVis}.
An initially weak fluctuation of $\Pi_l$ at level $l=4$ transfers energy to the next level, $l=5$, causing a stronger fluctuation that is shifted in time and lasts for a shorter time.
This process continues as long as viscosity remains ineffective.
In the inviscid problem, such sequences of energy-flux pulses reach $l\to\infty$ in finite time and correspond to singular solutions first calculated in~\cite{dombre1998intermittency}.
This mechanism, in which the amplitude of $\Pi_l$ is amplified locally in time, has been investigated extensively in~\cite{mailybaev2013,mailybaev2012computation,mailybaev2012renormalization}.
%\ALX{The same dynamical amplification mechanism is present in the multi-branch case, but each pulse can now feed more than one descendant branch.}
\medskip

In simple terms, this amplification can be viewed as follows.
Fluctuations at the large scales appear intrinsically due to the chaotic dynamics and pass down to smaller scales through nonlinear processes.
A fluctuation of amplitude $\Pi_l$ at scale $\ell_l={k_l}^{-1}$ with duration $\tau_l$ will transfer part of its energy $E_l\propto \Pi_l\tau_l$ to scale $\ell_{l+1}$, generating a new fluctuation at that scale.
Here, we consider the local maximum in time of $\Pi_l(t)$ as the amplitude of the fluctuation.
On physical ground~\cite{eyink2005, alexakis2018cascades}, we can safely assume (i) the locality in scale of the interactions and (ii) the inviscid dynamics of the inertial range; then the duration of such a pulse can only depend on the scale $\ell_l$ and the amplitude $\Pi_l$.
Therefore, for dimensional reasons, it has to scale as
\begin{equation}
    \tau_l \propto {\Pi_l}^{-1/3} \,{\ell_l}^{2/3}. 
    \label{eq:time}
\end{equation}
If a fraction $\chi$ of the energy is passed from one scale to the next, then the energy at scale $l$ is
\begin{equation}
    E_l\propto \chi^l E_0, \qquad E_l\propto \Pi_l \tau_l \propto (\Pi_l \ell_l)^{2/3},
    \label{eqEl}
\end{equation}
where Eq.~\eqref{eq:time} has been used.
Therefore, the amplitude of the energy-flux pulse and its duration are, respectively,
\begin{align}
     \Pi_l &\propto {E_0}^{3/2}{\ell_0}^{-1} (\chi^{3/2}\lambda)^{l},
     \label{eq_Pi_l_scaling}
     \\ 
     \tau_l &\propto {E_0}^{-1/2}\ell_0 (\chi^{1/2}\lambda)^{-l},
     \label{eq_tau_el_scaling}
\end{align}
where we have used Eq.~\eqref{eqEl} and $\ell_l=\ell_0\lambda^{-l}$.
Equation~\eqref{eq_Pi_l_scaling} predicts that if $\chi^{3/2}\lambda>1$, then $\Pi_l$ grows with $l$, as pointed out by results shown in Fig.~\ref{figVis}.
The ratio, however,
\begin{equation}
    \pi_l = \Pi_l/\Pi_{l-1}
    \label{eqmult}
\end{equation}
is $l$-independent and equal to $\pi_l = \chi^{3/2}\lambda$.
The ratios in Eq.~\eqref{eqmult} are referred to as multipliers because the flux amplitude at scale $l$ can be written as their product, 
\begin{equation}
    \Pi_l=\Pi_0 \prod_{j=1}^l \pi_j,
    \label{product}
\end{equation}
where $\Pi_0$ is the flux at level $l=0$.
%The same dynamical amplification mechanism is present in the multi-branch case, but each pulse can now feed more than one descendant branch.
%
In the particular case $\chi=1$, all the energy is transferred from one pulse to the next.
In this case, if relations~\eqref{eqEl} and~\eqref{eq_Pi_l_scaling} were exact and if all pulses were generated only at large scales, one would obtain a monofractal behavior, different from the Kolmogorov one, with $\langle {\Pi_l}^m\rangle\propto \lambda^{(m-1)l}$.
However, this is not what is observed in Fig.~\ref{figScaling}. \medskip

There are two reasons why such a monofractal behavior is not observed.
First, not all energy is transferred from one scale to the next.
In~\cite{mailybaev2013}, $\chi$ was found to be $0.68$ for instanton solutions, whose amplitude peaks follow a power law $|u_l|\propto {\ell_l}^{y_0}$, giving $\chi=\lambda^{-2y_0}$ by dimensional analysis.
Numerical simulations of the turbulent Sabra shell model give $\chi=0.74$.
This implies that only a fraction of the energy transferred from level $l-1$ to level $l$ is passed down to level $l+1$, while the remaining energy is stored at level $l$.
This energy will also need to cascade at later times through the generation of new pulses.
Energy conservation thus implies that new peaks, i.e., new local maxima, are generated at every level $l$.
\begin{figure}[h]
\includegraphics[width=.9\columnwidth]{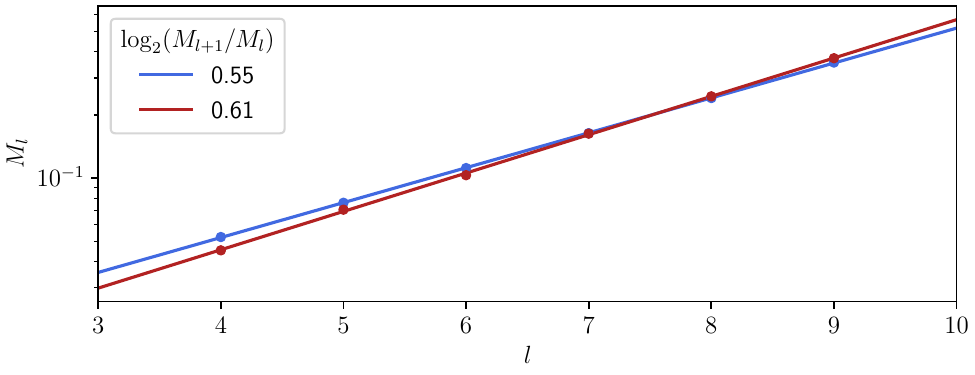}
\caption{\label{figNumPeaks} Normalized number of peaks as a function of $l$ for the single-branch model (blue) and the multi-branch model (red), with the latter averaged over all nodes at fixed level.}
\end{figure}
Figure~\ref{figNumPeaks} shows the scaling of the number of maxima $M_l$ at level $l$, for both the single-branch and the multi-branch models.
It grows exponentially with $l$, equivalently as a power law in $\ell_l$,
\begin{eqnarray}
    M_l \propto 2^{\alpha l} = {\ell_l}^{-\alpha\log 2/\log\lambda }
\end{eqnarray}
with $\alpha=0.55$ for the single-branch model.
This means that, approximately for every two peaks crossing a given level $l$, a third, generally weaker, peak is self-generated by chaotic fluctuations and is not necessarily causally related to a peak formed at previous levels.
Examples of such new peaks can be seen at the largest values of $l$ in Fig.~\ref{figFlux}.\medskip

A similar picture holds for the $2$-adic model.
The right panel of Fig.~\ref{figVis} shows $\Pi_{l,n}$ for $l=4$ and $n=1$ with a black solid line.
The energy fluxes of its two descendants are shown in brown, using a solid line for the larger one and a dashed line for the smaller one.
We continue in this way, showing only the descendants of the larger branch (keeping all descendants would already produce $32$ lines at level $l=9$).
A pulse at scale $l$ generates two pulses, one for each descendant at scale $l+1$, whose normalized amplitudes are larger and whose durations are shorter, although the two pulses are not necessarily equal.
In Fig.~\ref{figNumPeaks}, we also plot the scaling of the number of peaks $M_l$, averaged over $n$.
The number of maxima at each level also increases with $l$, at a slightly higher rate than in the single-branch case.
We observe $\alpha=0.61$ for the multi-branch model, compared to $\alpha=0.55$ for the single-branch.
This provides a comparison of the dynamical source of intermittency, since spatial configurations are averaged out.
The increase of $\alpha$ suggests that the peak-production mechanism is enhanced by the presence of a nontrivial geometrical structure.\medskip

The second reason why monofractality is not observed is that, due to the omnipresent chaotic fluctuations in the system, relations~\eqref{eq:time} and~\eqref{eq_Pi_l_scaling} are not exact but fluctuate depending on the instantaneous noise level.
This implies that the multipliers in Eq.~\eqref{eqmult} do not take a unique value, but rather fluctuate randomly according to a distribution.
In the next section, we compute this distribution and assess whether it is scale invariant, namely independent of the level $l$, as suggested by Eq.~\eqref{eqmult}.

% --------------------------------------------------------------------------- |

\subsection{Multipliers}

\begin{figure}[h]
\includegraphics[width=.9\columnwidth]{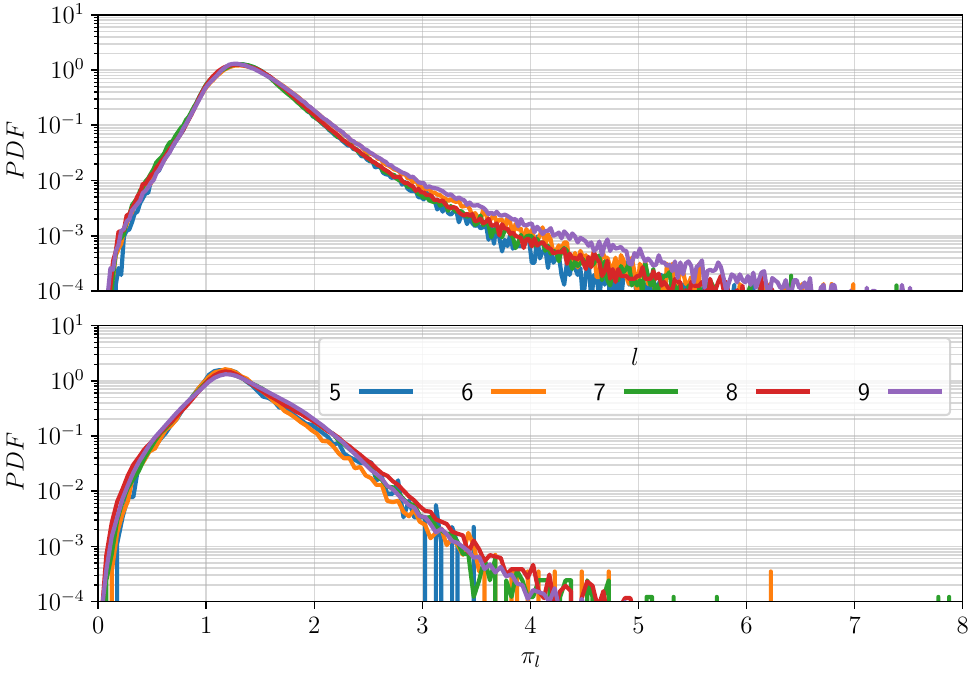}
\caption{\label{figRatios} Flux-multiplier probability distributions. Classical $1$-adic model (top) and multi-branch $2$-adic model (bottom). For the multi-branch model, different spatial samples are collected as realizations of the same random variable.}
\end{figure}
We describe the procedure used to compute flux multipliers in the multi-branch setting, which includes the single-branch case.
For the dyadic model,
\begin{equation}
    \pi_{l,n}=\Pi_{l,n}/\Pi_{l-1, \lceil n/2\rceil}.
\end{equation}
We then identify pairs of flux maxima whose ratio defines the relevant multiplier through the following procedure.
First, for each node, we consider all maxima of $\Pi_{l,n}$ that exceed the mean energy injection rate $\epsilon$. We then relate maxima according to temporal causality and the genealogical constraint. 
Let us make an example: let us assume that, for the node $(l,n)$, two consecutive maxima above $\epsilon$ occur at times $t_1$ and $t_2$.
All maxima of the children, $(l+1,2n-1)$ and $(l+1,2n)$, occurring in the time interval $(t_1,t_2)$ are potentially related to the maximum at $t_1$.
We consider only the first occurring ratio for each child, since the probability of having more than one related maximum in the next shell is small, and additional maxima are not necessarily causally related to the peak at the previous level. \medskip

Starting from the single-branch data, Fig.~\ref{figRatios} shows the distributions of $\pi_l$ for various values of $l$ in the inertial range.
Contrary to the PDFs of $\Pi_l$, which do not follow a self-similar behavior, the distributions of the multipliers $\pi_l$ overlap for different $l$, implying a scale-invariant behavior.
The most important point implied by these results is that, given the state at level $l-1$ in the inertial range, 
the statistical behavior at level $l$ is predictable and independent of the value of $l$.
In other words the local dynamical behavior is scale invariant up to an amplitude rescaling. 
%\SC{(I am not satisfied by this sentence. Let discuss it Flavio.)}
\medskip

The multipliers $\pi_{l,n}$ follow distributions similar to the single-branch ones.
In particular, the distributions collapse in $l$, implying a scale invariant behavior of the multipliers.
The PDF shows a peak at the same value as in the single-branch case. % and slightly weaker tails.
%Despite the weaker tails, 
However, we have seen that for the multi-branch model there is still an enhancement of intermittency in Fig.~\ref{figScaling}.
The explanation has to be related to the uneven splitting of the energy flux between the two descendants, which is absent in single-branch models by construction.
%\SC{(correct?)}
%
\begin{figure}[h]
\includegraphics[width=.9\columnwidth]{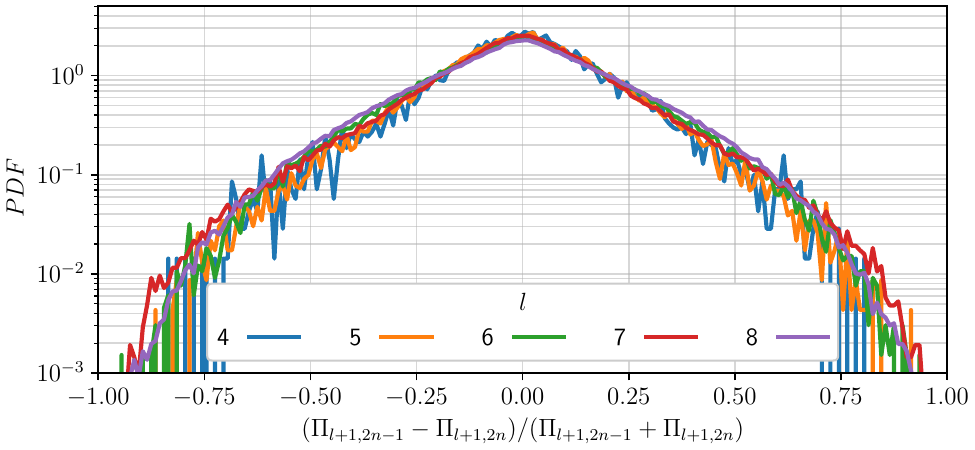}
\caption{\label{figAsymm} Ratio measuring the partition of the energy flux between brothers in the dyadic model.}
\end{figure}
We quantify this source of intermittency by looking at the PDFs of the ratio
\begin{equation}
    A= 
    \frac{\Pi_{l,n}-\Pi_{l,n+1}}{\Pi_{l,n}+\Pi_{l,n+1}} 
    =\frac{\pi_{l,n}-\pi_{l,n+1}}{\pi_{l,n}+\pi_{l,n+1}} 
\end{equation}
where $(l,n)$ and $(l,n+1)$ are two brother nodes.
The value $A=0$ implies a perfectly even distribution of the cascaded energy between the two brothers, while $A=\pm1$ implies that only one of the two brothers receives the energy, whereas the other and all its descendants receive no energy. The computed distribution is plotted in Fig.~\ref{figAsymm}.
The PDF is significantly spread, with half of the events having $|A|>0.13$.
This implies that variations of the flux across space, i.e., across $n$, increase rapidly as larger values of $l$ are reached.
\medskip

Treating the multipliers as random variables implies that the energy flux at scale $l$ can be written as a multiplicative random process, whose distribution can be calculated using large deviation theory~\cite{frisch1995turbulence,ellis1984large}.
This description was used in~\cite{mailybaev2013}, successfully reproducing the multifractal behavior.
We directly test the large deviation theory in the final section.

% --------------------------------------------------------------------------- |

\section{Large Deviation Theory}

The results of the previous section indicate that the flux statistics are governed by an underlying multiplicative process. In the single-branch model, the energy flux at level $l$ is written as the product of the multipliers $\pi_l$, as described by Eq.~\eqref{product}. Equivalently, in the multi-branch model, one has
\begin{equation}
    \Pi_{l,n} = \Pi_0 \prod_{j=0}^{l-1} \pi_{l-j,\lceil n/2^j\rceil}.
\end{equation} 
The logarithm of $\Pi_l$, or of $\Pi_{l,n}$ in the multi-branch model, can then be written as a sum of log multipliers, with $r_l=\log \pi_l$ and $r_{l,n}=\log \pi_{l,n}$, respectively. 
For the multi-branch model, along the cascade path ending at the node $(l,n)$, we define
\begin{equation}
    Y_{l,n}=
    \frac{1}{l}\log\frac{\Pi_{l,n}}{\Pi_0}=
    \frac{1}{l}\sum_{j=0}^{l-1}r_{l-j,\lceil n/2^j\rceil}.
\end{equation}
An analogous definition holds in the single-branch model, with the sum taken along the unique cascade path. Thus, $Y_{l,n}$ is a finite scale average of the logarithmic multipliers.
If the multiplier statistics become approximately stationary along the cascade, the typical fluctuations of $Y_{l,n}$ are expected to concentrate around their mean as $l$ increases. However, the flux itself depends exponentially on this average,
%\begin{equation}
$\Pi_{l,n}=\Pi_0 \exp[lY_{l,n}]$.
%\end{equation}
As a consequence, high order moments of the flux are sensitive to rare fluctuations of $Y_{l,n}$ rather than only to its typical, Gaussian fluctuations. This is precisely the regime in which large deviation theory provides the appropriate description.
\medskip
 
The Gärtner-Ellis theorem~\cite{ellis1984large,touchette2009large} states that, if the sequence $Y_l$ satisfies a large deviation principle, its probability density has the asymptotic form
\begin{equation}
    p_l(y) \underset{l\,\to\,\infty}{\asymp} \exp[-l I(y)],
\end{equation}
where $I(y)$ is the large deviation, or Cramér, rate function.
The rate function can be computed analytically if the distribution of the logarithmic multipliers $r_l$ is known.
Here, we directly calculate $I$ from $p(y)$ 
by defining the finite-$l$ estimator $I_l(y)$ as   
\begin{eqnarray}
    I_l(y) = -\frac{1}{l}\log p_l(y)
\end{eqnarray}
and demonstrating that $I_l(y)$ tends to an $l$ independent shape as $l\to\infty$.
\begin{figure}[h]
\includegraphics[width=.9\columnwidth]{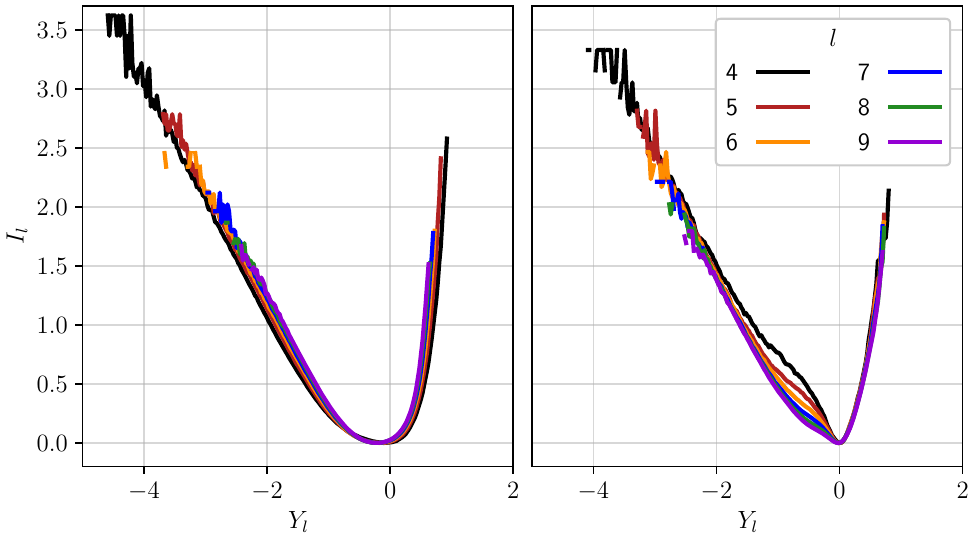}
\caption{\label{figRateFunc} Rate functions for the energy flux process. Single-branch model on the left and multi-branch model on the right, with the latter averaged in space in addition to time.}
\end{figure}
This is shown in Fig.~\ref{figRateFunc} for the single-branch and multi-branch cases. The left panel refers to the single-branch model and displays a good collapse for levels $l$ in the inertial range. A vertical shift has been applied to compensate for normalization differences between the estimated PDFs. We stress that it is the large values of $y$ that determine intermittency, which corresponds to extreme positive fluctuations of the flux.
The right panel shows the analogous estimate for the multi-branch model.
To build the PDFs of the logarithmic variables, we use different spatial realizations, i.e., different values of $n$ at fixed $l$, as samples of the same distribution.
Since the number of spatial samples grows exponentially with $l$, the largest levels provide better statistics. 
Overall, these results support a large deviation description of the multiplicative energy cascade.
This is remarkable because flux multipliers across levels are not independent, whereas independence is assumed in the simplest setting of Cramér's theorem~\cite{ellis1984large,touchette2009large}.

% --------------------------------------------------------------------------- |

\section{Conclusions}

We have analyzed in detail the results from simulations of two turbulence shell models, focusing on the intermittent behavior of the energy flux towards small scales.
We have shown that a dynamically evolving multi-branch shell model displays stronger intermittency than the corresponding single-branch Sabra model.
This is seen directly in the energy-flux structure functions, whose scaling exponents deviate more strongly from the scale invariant prediction.\medskip

The analysis of the flux signals separates two contributions to intermittency.
The first is the dynamical amplification already present in the single-branch model: flux bursts become stronger and shorter as they propagate toward smaller scales.
This mechanism is associated with the multiplier statistics of the flux amplitudes.
The second contribution is specific to the multi-branch model: the energy flux is split unevenly between brother nodes.
As a consequence, even when the space-averaged flux remains weakly fluctuating, the local flux at individual nodes develops strong spatial variations.\medskip

This interpretation is supported by the statistics of peaks and multipliers.
The number of flux local maxima grows exponentially with the level, showing that new transfer events are generated along the cascade.
The growth rate is slightly larger in the multi-branch model, suggesting that the geometrical structure enhances the production of peaks.
At the same time, the multiplier distributions collapse across levels, indicating scale invariant local transfer statistics.
The asymmetry ratio between brother nodes further shows that a significant fraction of events corresponds to an uneven partition of the energy flux.\medskip

Finally, the rate-function analysis supports a large deviation description of the multiplicative energy-flux process.
The collapse of the finite-level estimators is observed both in the single-branch and in the multi-branch case, despite the fact that flux multipliers are not independent across levels.
This suggests that large deviation theory provides a useful effective description of the cascade statistics beyond the simplest independent-multiplier picture.
This result might also provide support to the proposals to deal with the cascade process in terms of some Markov processes~\cite{peinke2019fokker}.\medskip

Overall, the results show that the multi-branch model retains the dynamical intermittency of classical shell models while adding a minimal geometrical mechanism for spatial intermittency.
This makes it a useful intermediate setting between classical shell models and Navier-Stokes turbulence.
A natural next step is to test whether the same separation between dynamical amplification and spatial flux partition can be identified in direct numerical simulations of the Navier-Stokes equations. We plan to address this issue in our future work.

% --------------------------------------------------------------------------- |

\begin{acknowledgments}
This work received financial support from the CNRS through the MITI interdisciplinary initiatives, under its exploratory research program.
SC acknowledges funding from the ANR SCALP (ANR-24-CE23-1320).
This work was granted access to the HPC resources of GENCI-TGCC and GENCI-CINES under Project Nos. A0190506421, A0172B10759, and AD010517900.
\end{acknowledgments}

% --------------------------------------------------------------------------- |

% \nocite{*}

\bibliography{apssamp}

\end{document}